########################################################################################

\documentstyle[12pt]{article}
\newcommand{\rank}{\mathop{\rm rank}\nolimits}
\newcommand{\diag}{\mathop{\rm diag}\nolimits}

\begin{document} \large

\date{}
\title{\hspace*{13.5cm}{\large USITP-98-23}\\
\vspace*{2cm}$N=2$ massive superparticle:\\
the minimality principle and the k-symmetry}
\author{D.V.Uvarov$^{a}$ \thanks{E-mail:uvarov@nik.kharkov.ua}\, \,and
 A.A.Zheltukhin$^{a,b}$ \thanks{E-mail:aaz@physto.se}\\
{\normalsize $^{a}$ NSC Kharkov Institute of Physics and Technology}\\
{\normalsize 310108, Kharkov, Ukraine,}\\
{\normalsize $^{b}$ Institute of  Theoretical Physics , University of  Stockholm}\\
{\normalsize Box 6730, S-11385 Stockholm, Sweden}}
\maketitle
\begin{abstract}
The electromagnetic interaction of massive superparticles with $N=2$ extended Maxwell
supermultiplet is studied. It is proved that the minimal coupling breaks  the k-symmetry.
A non-minimal k-symmetric action is built and it is established that the k-symmetry
uniquely fixes the value of the superparticle's anomalous magnetic moment.
\end{abstract}

 \vspace*{0.5cm}\begin{center}{\Large\bf 1. Introduction}\end{center}
\vspace*{0.5cm}

 It is well known that consistency of the theories of superstrings,
super-$p$~ -(and $D$-) branes [1,2] may be achieved only if the k-symmetry is present in
these theories [3-5]. However, the presence of the k-symmetry in a free supersymmetric
theory doesn't guarantee its presense in a theory with interaction.
This fact accounts for an interest to the problem of the k-symmetry preservation when passing
from free supersymmetric models to ones with interaction. The model of a charged superparticle
coupled to external superpotential [7-9, 12] is one of the simplest supersymmetric schemes
 with interaction.
In the case of $N=1$ massless superparticle the model with interaction
possessing the k-symmetry can be obtained by strict following to the minimality priciple.
 An important consequence of  the k-symmetry existence is presence  of the correct constraints
 for superfield strengths which remove unphysical fields and single out the $N=1$ Maxwell
 supermultiplet. Analogous situation takes place  in the case of $N=1$ supergravity.

In Ref.[4] has been proposed an extended $(N>1)$ free superparticle model which possesses the k-symmetry and, unlike the $N=1$ case,
permits the  covariant (although reducible) division of the grassmanian constraints without
 introduction of the auxiliary variables such as
twistor-like ones [10]. However, consistency of the minimality principle
and the k-symmetry  is violated when passing to that superparticle model
coupled to extended $N>1$ superpotential. As revealed by Luzanna and Milevski in Ref.[12],
this breakdown requires, the modifications of the model, analogous to those of the Refs.[13,14] for the
spinning particles. These modifications, based  on the mass ``renormalization'' prescription,
turned out to be equivalent to the introduction of an anomalous magnetic moment
(AMM) for the spinning particles as it was established in Ref.[15], where a superfield description
 was given. Moreover, no restrictions on the AMM magnitude of the spinning particle
appeared.

In some sense a similar situation appears in the model of $N=2$ massive superparticle which is
studied below. However, here the requrement of the k-symmetry existence severely restricts
the AMM value of superparticle. Clarification of this statement is a reason for this paper.

The paper is organised as follows: section 2 is devoted to the study of the free $N=2$ massive
superparticle model [4]. In the section 3 we examine this model when the minimal coupling
with an external $N=2$ superpotential is introduced. Then the problem of the k-symmetry
breaking for this coupling is discussed.
In the section 4 we build  k-symmetric action for the superparticle interacting with
the external $N=2$ superpotential. We show that the restoration of the k-symmetry of the
action is provided by means of the nonminimal terms introduction. This nonminimal extension of
the model corresponds to taking into account the electromagnetic interactions of the
superparticle caused by its AMM. Moreover, the value of this AMM turns out to be fixed.

\begin{center}
{\Large\bf 2.  $N=2$ massive superparticle model}
\end{center}

Among various superparticle models of the most interest  are those,
which possess the k-symmetry. In particular, these are massless superparticles in $D=3,4,6,10$
with an arbitrary number of supersymmetries [5]. At the same time a transition to the corresponding massive superparticle  model violates the k-symmetry.
However, when $N>1$ there exists the possibility to avoid these difficulties by means of Wess-Zumino -Witten-like term introduction.
For the  $D=4$ case such model was suggested in [4] and for the $D=6,10$ in [6].

We resort the $D=4$, $N>1$ case [4], where to the Brink-Schwarz action was added the additional
 term
$${\displaystyle \theta^{\alpha }_iA^{ij}\dot{\theta }_{\alpha  j}+
\bar{\theta }_{\dot{\alpha }i}A^{ij}\dot{\bar\theta }\mathstrut^{\dot\alpha }_j,}$$
where $A^{ij}$ is a real antisymmetric matrix depending on the superparticle's mass.
This term is invariant  under global  supersymmetry transformations
up to the total derivative and is a  1D analogue of the super-p-branes
Wess-Zumino-Witten term [5]. In the $N=2$ case this matrix is simply
$m\epsilon ^{ij}$  and the  superparticle action takes form
\begin{equation}
{\displaystyle
S=\int\! d\tau\sqrt{-\omega^{\mu}\omega_{\mu}}
+m\int\! d\tau\left(\theta ^{\alpha }_i\dot\theta ^i_{\alpha }
+\bar{\theta}_{\dot\alpha  i}                 \dot{\bar\theta}\mathstrut^{\dot\alpha  i}\right),}
\end{equation}
where $\omega ^{\mu }=\dot x^{\mu }+i\theta ^{\alpha }_i\sigma ^{\mu }_{\alpha
\dot{\alpha }}\dot{\bar\theta }\mathstrut^{\dot\alpha  i}-\dot\theta ^{\alpha }_i
\sigma ^{\mu }_{\alpha  \dot\alpha }\bar\theta ^{\dot\alpha  i}$
are the supersymmetric Cartan forms. Our notations mainly coincide with those
of the Ref.[11] (see also Appendix A). Introducing the worldline einbein g action (1) can be represented in the
following form
$$
{\displaystyle
S=\frac12\int\! d\tau\left(\frac{\omega^2}{g}-gm^2\right)+
m\int\! d\tau\left(\theta^{\alpha}_i\dot\theta^i_{\alpha}+\bar\theta_{\dot\alpha i}
\dot{\bar\theta}\mathstrut^{\dot\alpha i}\right).
}\eqno (1')
$$

The Hamiltonian analysis [16] we begin with the introduction of the momentum variables
\begin{equation}
\begin{array}{rl}{\displaystyle
p_{\mu }=\frac{\partial L}{\partial\dot x^{\mu }}=\frac{\omega_{\mu }}{g},}&
{\displaystyle p_g=\frac{\partial L}{\partial\dot g}=0}\\
[0.3cm]
{\displaystyle \pi^i_{\alpha }=\frac{\partial L}{\partial{\dot\theta^{\alpha }_i}}=
-i\frac{\bar\theta ^{\dot\alpha  i}\omega_{\alpha \dot\alpha }}{g}                                       -m\theta ^i_{\alpha },}&
{\displaystyle \bar\pi _{\dot\alpha  i}=\frac{\partial L}{\partial\dot{\bar\theta }\mathstrut^{\dot\alpha  i}}=-i\frac{\theta ^{\alpha }_i\omega _{\alpha \dot\alpha }}{g}
-m\bar\theta_{\dot\alpha  i},}
\end{array}\end{equation}
leading to the set of the primary constraints $p_g\approx 0$,
\begin{equation}
V^i_{\alpha}=\pi ^i_{\alpha}+ip_{\alpha \dot\alpha}\bar\theta ^{\dot\alpha i}+
m\theta ^i_{\alpha}\approx 0; \hspace{1em}                                                                 \bar V_{\dot\alpha i}=\bar\pi _{\dot\alpha i}+
i\theta ^{\alpha}_ip_{\alpha \dot\alpha }+m\bar\theta_{\dot\alpha i}\approx 0
\end{equation}
and the standard  Hamiltonian
\begin{equation}{\displaystyle
H_0=\dot x^{\mu }p_{\mu }+\dot\theta ^{\alpha}_i\pi ^i_{\alpha}+
\dot{\bar\theta }\mathstrut_{\dot\alpha i}\bar\pi ^{\dot\alpha i}-L=
\frac{g}{2}\left(p^2+m^2\right).}
\end{equation}
Having the constraints requires introduction of the full Hamiltonian
$$
{\displaystyle
H=H_0+\lambda^{\underline\alpha }V_{\underline\alpha }+
\bar\lambda_{\underline{\dot\alpha }}\bar V^{\underline{\dot\alpha }}+\varphi p_g,}
\eqno (4')
$$
which is a linear combination of the primary constraints. To proceed further we
need to have the Poisson brackets definition
\begin{equation}
\begin{array}{rl}{\displaystyle
\left\{p_{\mu },x^{\nu }\right\}=-i\delta_{\mu }^{\nu },}&
{\displaystyle
\left\{p_g,g\right\}=-i,}\\[0.4cm]
{\displaystyle
\left\{\pi_{\alpha }^i,\theta_j^{\beta }\right\}=-i\delta_j^i\delta_{\alpha }^{\beta },}&
{\displaystyle
\left\{\bar\pi_{\dot\alpha i},\bar\theta^{\dot\beta j}\right\}=-i\delta_i^j
\delta^{\dot\beta}_{\dot\alpha}.}
\end{array}
\end{equation}
Using (5)  we can evaluate several important Poisson brackets
\begin{equation}{\displaystyle
\left\{V_{\underline\alpha },V_{\underline\beta }\right\}=-2im
\epsilon_{\underline\alpha\underline\beta},
\left\{\bar V_{\underline{\dot\alpha}},\bar V_{\underline{\dot\beta}}\right\}=
-2im\epsilon_{\underline{\dot\alpha}\underline{\dot\beta}},
\left\{V_{\underline\alpha },\bar V_{\underline{\dot\alpha}}\right\}=
2p_{\underline\alpha\underline{\dot\alpha }},}
\end{equation}
\begin{equation}{\displaystyle
\left\{p^2+m^2,V_{\alpha }\right\}=0,\hspace{1em}
\left\{p^2+m^2,\bar V\right\}=0.}
\end{equation}
Now, according to the Dirac prescription [16], we study equations, obtained from the primary
 constraints by the temporal conservation conditions:
\begin{equation}{\displaystyle
\left\{p_g,H\right\}=0\Longrightarrow \chi=p^2+m^2\approx0,
\mbox{(secondary constraint)}}
\end{equation}
\begin{equation}{\displaystyle
\left\{H,V^i_{\alpha }\right\}=-2im\lambda^i_{\alpha}+2p_{\alpha\dot\beta}
\bar\lambda^{\dot\beta i}=0,}\hspace{1em}
{\displaystyle
\left\{H,\bar V_{\dot\alpha i}\right\}=2\lambda^{\beta}_ip_{\beta\dot\alpha}-
2im\bar\lambda_{\dot\alpha i}=0.}
\end{equation}
A system of linear equations has nontrivial solutions when and only when its determinant vanishes. In our case
\begin{equation}{\displaystyle
\det A\equiv\det\left(
\begin{array}{cc}{\displaystyle
-im\delta^{\beta}_{\alpha}\delta^i_j}& {\displaystyle
\delta^i_jp_{\alpha\dot\beta}}\\[0.4cm]
{\displaystyle -\delta^i_j\tilde p^{\dot\alpha\beta}}&
{\displaystyle -im\delta^{\dot\alpha}_{\dot\beta}\delta^i_j}
\end{array}
\right)={\left(p^2+m^2\right)}^4\approx0.}
\end{equation}
This condition however does not fix system's rank, which in our case equals four
\footnote{\normalsize Unlike the interaction case here we need no additional conditions to
 halve the rank to obtain the k-symmetry}, so only half of the equations (9) are linear independent and, as a consequence, four of the Lagrange multipliers $\lambda(\bar\lambda)$ remain unfixed. This indicates the  presence of the local fermionic symmetry of the action $(1')$, which is
 just the
k-symmetry, and existence of the four spinorial first-class constraints, generating this
 symmetry. The rest of the spinorial constraints in (3) belong to the second-class.
 At this point the new problem arises: how is it possible to produce the covariant division
 of the first- and second-class spinorial constraints? The matrix A (10) prompts us to use
the set of projectors [17], separating the first- and second-class constraints
\begin{equation}{\displaystyle
P_{I,II}=\frac12\left(1\pm\Pi\right),}
\end{equation}
where
$$                                                                                                                                                                     {\displaystyle
\Pi=\delta^i_j\left(
\begin{array}{cc}
{\displaystyle 0}& {\displaystyle -i\frac{p_{\alpha\dot\beta}}{\sqrt{-p^2}}}\\[0.4cm]
{\displaystyle i\frac{\tilde p^{\dot\alpha \beta}}{\sqrt{-p^2}}}&
{\displaystyle 0}
\end{array}
\right)}
\eqno (11')
$$
and $\Pi$ satisfies the  strong relation $\Pi^2=1$. The multiplier $\left(\sqrt{-p^2}\right)$ was introduced here for a normalization. The projectors $P_I$ and $P_{II}$ obey the
 following relations: $P^2_{I,II}=P_{I,II};{} P_{I}P_{II}=P_{II}P_{I}=0$. Then the first-and
second-class constraints acquire the form
\begin{equation}
\begin{array}{c}
\begin{array}{rcl}
{\displaystyle\left(
\begin{array}{c}
{\displaystyle V^{(1)}{}^i_{\alpha}}\\{\displaystyle\bar V^{(1)\dot\alpha i}}
\end{array}
\right)}& =P_{I}\left(
\begin{array}{c}
{\displaystyle V^j_{\beta}}\\{\displaystyle\bar V^{\dot\beta j}}
\end{array}\right)& =\frac12\left(
\begin{array}{c}
{\displaystyle V^i_{\alpha}-i\frac{p_{\alpha\dot\beta}\bar V^{\dot\beta i}}{\sqrt{-p^2}}}
\\
{\displaystyle \bar V^{\dot\alpha i}+i\frac{\tilde p^{\dot\alpha\beta}V^i_{\beta }}
{\sqrt{-p^2}}}
\end{array}\right),
\end{array}\\[1.0cm]
\begin{array}{rcl}
{\displaystyle\left(
\begin{array}{c}
{\displaystyle V^{(2)}{}^i_{{}\alpha}}\\{\displaystyle\bar V^{(2)\dot\alpha i}}
\end{array}\right)}&
=P_{II}\left(\begin{array}{c}{\displaystyle V^j_{\beta}}\\
{\displaystyle\bar V^{\dot\beta i}}
\end{array}\right)& =\frac12\left(
\begin{array}{c}
{\displaystyle V^i_{\alpha}+i\frac{p_{\alpha\dot\beta}\bar V^{\dot\beta i}}           {\sqrt{-p^2}}}\\
{\displaystyle\bar V^{\dot\alpha i}-i\frac{\tilde p^{\dot\alpha\beta}V^i_{\beta}}
{\sqrt{-p^2}}}
\end{array}\right).
\end{array}
\end{array}
\end{equation}
Although we managed to separate constraints in the manifestly covariant way we have got,
however, the
 linearly dependent sets of the first- and second-class constraints
\begin{equation}
{\displaystyle i\frac{\tilde p^{\dot\beta\alpha}V^{(1)}{}_{\alpha}^i}
{\sqrt{-p^2}}=\bar V^{(1)\dot\beta i},\hspace{2em}
-i\frac{\tilde p^{\dot\beta\alpha}V^{(2)}{}_{\alpha}^i}{\sqrt{-p^2}}=
\bar V^{(2)\dot\beta i}.}
\end{equation}
One can use projector $P_{I}$ for constructing the k-symmetry transformation laws for
the action (1): ${\displaystyle\delta x^{\mu}=                                 -i\theta_i\sigma^{\mu}\delta\bar\theta^i
+i\delta\theta_i\sigma^{\mu}\bar\theta^i,}$
\begin{equation}
\begin{array}{rcl}
\left(\begin{array}{c}{\displaystyle\delta\theta^i_{\alpha}}\\
{\displaystyle\delta\bar\theta^{\dot\alpha i}}
\end{array}\right)&
=P_{I}\left(\begin{array}{c}{\displaystyle\kappa^j_{\beta}}\\
{\displaystyle\bar\kappa^{\dot\beta j}}
\end{array}\right)&
=\frac12\left(\begin{array}{c}
{\displaystyle\kappa^i_{\alpha}-i\frac{\omega_{\alpha\dot\beta}
\bar\kappa^{\dot\beta i}}{\sqrt{-\omega^2}}}\\
{\displaystyle\bar\kappa^{\dot\alpha i} +i\frac{\tilde\omega^{\dot\alpha\beta}\kappa^i_{\beta}}
{\sqrt{-\omega^2}}}
\end{array}\right).
\end{array}
\end{equation}
The bosonic constraint $\chi$ belongs to the first-class here (being a reparametrization
generator), but the external superpotential coupling
converts it to the second-class as will be seen below
\begin{equation}
{\displaystyle\{\chi,\chi\}=0,\hspace{2em}  \{\chi,V_{\underline\alpha}\}=0,\hspace{2em}
\{\chi,\bar V_{\underline{\dot\alpha}}\}=0.}
\end{equation}

Thus, our analysis explicitly shows how to covariantly separate the fermionic constraints and
 to construct the k-symmetry transformations. The  total constraints algebra is presented in
the Appendix B.

The  next step in our analysis will be investigation of the consistency between  the k-symmetry
and the minimal coupling procedure to introduce the interaction of a superparticle with
an external superpotential.
\begin{center}
{\Large\bf3. $N=2$ massive superparticle\\ coupled to external superpotential}
\end{center}

We start from the following action of the $N=2$ massive charged superparticle coupled to an
 external superpotential
\begin{equation}
\begin{array}{l}{\displaystyle
S^{(e)}_{min}=\int d\tau\left[\frac12
\left(\frac{\omega^{\mu}\omega_{\mu}}{g}-
gm^2\right)+m\left(\theta^{\alpha}_i\dot\theta^i_{\alpha}+
\bar\theta_{\dot\alpha i}\dot{\bar\theta}\mathstrut^{\dot\alpha i}
\right)\right]
}\\[0.4cm]\hspace{3em}{\displaystyle
{}+ie\int d\tau\left(\omega^{\mu}A_{\mu}+\dot\theta^{\alpha}_i
A^i_{\alpha}+\dot{\bar\theta}\mathstrut_{\dot\alpha i}
\bar A^{\dot\alpha i}\right).
}\end{array}
\end{equation}
Here we restrict ourselves by the electromagnetic U(1) group case. The gauge superfields
 $A_{M}(x^{\mu},\theta,\bar\theta)=
\left(A_{\mu},A^i_{\alpha},A_{\dot\alpha i}\right)$ contain a great number of unphysical
component fields, which have to be removed by imposing gauge invariant  constraints
on the superfield strengths. The requirement of the k-symmetry existence will restrict the admissible form of these constraints.

 Now we consider the Hamiltonian treatment of the model (16) and introduce the canonical momenta
\begin{equation}
\begin{array}{c}
{\displaystyle p_{\mu}=\frac{\partial L}{\partial\dot x^{\mu}}=
\frac{\omega_{\mu}}{g}+ieA_{\mu},\hspace{2em} p_{g}=
\frac{\partial L}{\partial\dot g}=0}\\
{\displaystyle \pi^i_{\alpha}=\frac{\partial L} {\partial\dot\theta^{\alpha}_i}=-\frac{i\omega_{\alpha\dot\alpha}
\bar\theta^{\dot\alpha i}}{g}-m\theta^i_{\alpha}+
eA_{\alpha\dot\alpha}\bar\theta^{\dot\alpha i}+
ieA^i_{\alpha}}\\
{\displaystyle
\bar\pi_{\dot\alpha i}=
\frac{\partial L}{\partial\dot{\bar\theta}\mathstrut^{\dot\alpha }_i}
=-\frac{i\theta^{\alpha}_i\omega_{\alpha\dot\alpha}}{g}-
m\bar\theta_{\dot\alpha i}+
e\theta^{\alpha}_iA_{\alpha\dot\alpha}+ie\bar A_{\dot\alpha i},} \end{array}
\end{equation}
The primary constraints following from these definitions are the following: $p_{g}\approx0$
\begin{equation}
{\displaystyle V^i_{\alpha}=\pi^i_{\alpha}+ip_{\alpha\dot\alpha}
\bar\theta^{\dot\alpha i}+m\theta^i_{\alpha}-ieA^i_{\alpha}\approx0,
\bar V_{\dot\alpha i}=\bar\pi_{\dot\alpha i}+i\theta^{\alpha}_i
p_{\alpha\dot\alpha}+m\bar\theta_{\dot\alpha i}-
ie\bar A_{\dot\alpha i}\approx0,}
\end{equation}
and the canonical Hamiltonian is given by
\begin{equation}
{\displaystyle H_0=\frac{g}{2}\left[(p^{\mu}-ieA^{\mu})^2+m^2\right]}
\end{equation}
The total  Hamiltonian is
\begin{equation}
{\displaystyle H=H_0+\lambda^{\underline\alpha}
V_{\underline\alpha}+\bar\lambda_{\underline{\dot\alpha}}
\bar V^{\underline{\dot\alpha}}+\varphi p_g.}
\end{equation}
Below we remind some useful Poisson brackets relations
following from Eqs.(18-19)
\begin{equation}
\begin{array}{c}
{\displaystyle\{ V_{\underline\alpha},V_{\underline\beta}\}=-2im
\epsilon_{\underline{\alpha\beta}}-eF_{\underline{\alpha\beta}};
\hspace{2em}\{\bar V_{\underline{\dot\alpha}},
\bar V_{\underline{\dot\beta}}\}=-2im\epsilon_{\underline{\dot\alpha\dot\beta}}-
eF_{\underline{\dot\alpha\dot\beta}};}\\[0.4cm]
{\displaystyle\{ V_{\underline\alpha},\bar V_{\underline{\dot\alpha}}\}=
2{\cal P}_{\underline{\alpha\dot\alpha}}-
eF_{\underline{\alpha\dot\alpha}},\mbox{where }                                          {\cal P}_{\underline{\alpha\dot\alpha}}=
p_{\underline{\alpha\dot\alpha}}
-ieA_{\underline{\alpha\dot\alpha}};}\\[0.4cm]
{\displaystyle\{ H_0,V_{\underline\alpha}\}=\frac{eg}{2}
{\cal P}^{\dot\beta\beta}F_{\beta\dot\beta,\underline\alpha};
\hspace{2em}
\{ H_0,\bar V_{\underline{\dot\alpha}}\}=\frac{eg}{2}
{\cal P}^{\dot\beta\beta}F_{\beta\dot\beta,\underline{\dot\alpha}}.}
\end{array}
\end{equation}
The field strenghts are defined in [18]. Temporal conservation of the primary constraints
 leads to the secondary one $\chi={\cal P} +m^2\approx0$ and to the system of the linear
 equations for the Lagrange mulipliers
\begin{equation}
\begin{array}{c}
{\displaystyle E_{\underline\alpha}=Q_{\underline\alpha}+
\lambda^{\underline\beta}M_{\underline{\alpha\beta}}+
\bar\lambda^{\underline{\dot\beta}}N_{\underline{\alpha\dot\beta}}=0}
\\[0.4cm]{\displaystyle\bar E_{\underline{\dot\alpha}}=
\bar Q_{\underline{\dot\alpha}}+\lambda^{\underline\beta}
N_{\underline{\beta\dot\alpha}}+\bar\lambda^{\underline{\dot\beta}}
\bar M_{\underline{\dot\beta\dot\alpha}}=0}\\[0.4cm]
{\displaystyle Q^{\underline\alpha}\lambda_{\underline\alpha}+
\bar Q_{\underline{\dot\alpha}}\bar\lambda^{\underline{\dot\alpha}}=
0,}
\end{array}
\end{equation}
where we have  used the  following notations:
\begin{equation}
\begin{array}{ll}{\displaystyle
Q_{\underline{\alpha}}=\frac{eg}{2}{\cal P}^{\dot\beta\beta}
F_{\beta\dot\beta,\underline\alpha},}&{\displaystyle
\bar Q_{\underline{\dot\alpha}}=\frac{eg}{2}{\cal P}^{\dot\beta\beta}
F_{\beta\dot\beta,\underline{\dot\alpha}};}\\[0.4cm]{\displaystyle
M_{\underline{\beta\alpha}}=
-2im\varepsilon_{\underline{\beta\alpha}}-
eF_{\underline{\beta\alpha}},}&{\displaystyle
\bar M_{\underline{\dot\beta\dot\alpha}}=
-2im\varepsilon_{\underline{\dot\beta\dot\alpha}}-
eF_{\underline{\dot\beta\dot\alpha}};}
\end{array}
\end{equation}
$$
{\displaystyle
N_{\underline{\alpha\dot\alpha}}=
2{\cal P}_{\underline{\alpha\dot\alpha}}-
eF_{\underline{\alpha\dot\alpha}}.}
$$
It is easy to show that the last equation in (22) is a consequence of the others.
Similarly to the free case, the existence of the four spinorial first-class constraints
imposes certain restrictions on the rank of the system (22). Namely, this rank should be equal to four. The  matrix  of the system (22) equals
\begin{equation}
A=\left(\begin{array}{cc}
M& N\\[0.4cm]
N^T& \bar M
\end{array}
\right)
\end{equation}
Respectively, the matrix of the  extended system can be written in the form
\begin{equation}
A_{ex}=\left(
\begin{array}{ccc}
M& N& -Q\\[0.4cm]
N^T& \bar M& -\bar Q
\end{array}
\right).
\end{equation}
Using well-known properties of the rank invariance one can persuade that $\rank A=\rank R$ and
$\det A=\det R$, where
\begin{equation}
R=\left(\begin{array}{cc}
{\displaystyle M}& {\displaystyle N}\\[0.4cm]
0& {\displaystyle \bar M-N^TM^{-1}N}
\end{array}\right).
\end{equation}
We suppose that $\det M\not=0$, so $\rank M=4$. Indeed, $\det M\not=0$ when the interaction is
turned off, so  it is quite reasonable to assume  its conservation when interaction is
 turned on. Then we find that
 $\rank A=\rank M=\rank R=4$ and
\begin{equation}
{\displaystyle \bar Y_{\underline{\dot\beta\dot\alpha}}=                                                           \bar M_{\underline{\dot\beta\dot\alpha}}-
N_{\underline{\alpha\dot\beta}}M^{-1\underline{\alpha\beta}}
N_{\underline{\beta\dot\alpha}}=0.}
\end{equation}
 The system (22) is compatible, when and only when $\rank A=\rank A_{ex}=4$ or,
 equivalently,
\begin{equation}
\rank A_{ex}=rank\left(\begin{array}{ccc}
{\displaystyle M}& {\displaystyle N}& {\displaystyle -Q}\\[0.4cm]
0& 0& {\displaystyle -\bar Q+N^TM^{-1}Q}
\end{array}\right)=4,
\end{equation}
The latter equation implies the constraint
\begin{equation}{\displaystyle
\bar Q_{\underline{\dot\alpha}}-N_{\underline{\beta\dot\alpha}}
M^{-1\underline{\alpha\beta}}Q_{\underline{\alpha}}=0}
\end{equation}
provided that $\det N\not=0$.

Thus, we conclude that the system's rank is halved, if and only if the new constraints (27,29)
 are satisfied. In view of this observation the total Hamiltonian (20) should be extended
to the form
\begin{equation}
\begin{array}{l}
{\displaystyle H=\frac{g}{2}\chi-\frac12 Q^{\underline\alpha}M_{\underline{\alpha\beta}}
V^{\underline\beta}-\frac12\bar Q^{\underline{\dot\alpha}}
\bar M_{\underline{\dot\alpha\dot\beta}}\bar V^{\underline{\dot\beta}}+
\frac12\lambda^{\underline\alpha}\left(V_{\underline\alpha }-
N_{\underline{\alpha\dot\beta}}\bar M^{-1\underline{\dot\alpha\dot\beta}}
\bar V_{\underline{\dot\alpha}}\right)}\\[0.4cm]
\hspace{2em}{\displaystyle\frac12\bar\lambda^{\underline{\dot\alpha}}
\left(\bar V_{\underline{\dot\alpha}}-N_{\underline{\beta\dot\alpha}}
M^{-1\underline{\beta\alpha}}V_{\underline\alpha}\right)\approx 0,}
\end{array}
\end{equation}
where $\lambda^{\underline\alpha}$ and $\bar\lambda_{\underline{\dot\alpha}}$
are connected via either $E^1_\alpha,\bar E_{1\dot\alpha}$ or
$E^2_\alpha,\bar E_{2\dot\alpha}$\footnote{\normalsize As system's rank is halved now we can
consider any of these complex conjugate pairs as independent equations  .}.
The new Lagrange multipliers  $\lambda$  define the first-class constraints. It is well
 known that the first-class constraints form a closed algebra. To prove that we are dealing
with just this case, we are to calculate the Poisson brackets of the two second-class
 constraints
\begin{equation}
{\displaystyle\{ V^{(1)}_{\underline\alpha},V^{(1)}_{\underline\alpha}\}=
Y_{\underline{\alpha\beta}}+\mbox{(linear and quadratic terms in V($\bar V$))}\approx0,}
\end{equation}
where ${\displaystyle Y_{\underline{\alpha\beta}}=M_{\underline{\alpha\beta}}-
N_{\underline{\alpha\dot\beta}}\bar M^{-1\underline{\dot\beta\dot\alpha}}
N_{\underline{\beta\dot\alpha}}\approx0}$.\footnote{\normalsize Here and further  we
 omit explicit expressions for the terms proportional to the constraints V($\bar V$) because
they are irrelevant for the  definition of the  constraint's class.}
${\displaystyle Y_{\underline{\alpha\beta}}}$ is the complex conjugate constraint to
${\displaystyle\bar Y_{\underline{\dot\alpha\dot\beta}}}$. Note that it has the polynomial
structure in ${\displaystyle{\cal P}_{\underline{\alpha\dot\alpha}}}$
of the second power with the coefficients constructed from the spinorial components of the
 superfield strengths. This essential feature will play the crucial role in our further
analysis. Now we are going to study the Poisson brackets for
${\displaystyle V^{(1)}_{\underline\alpha}}$ and ${\displaystyle
Y_{\underline{\beta\gamma}}}$
\begin{equation}\begin{array}{l}
{\displaystyle \{ V^{(1)}_{\underline\gamma},
Y_{\underline{\alpha\beta}}\}=Y_{\underline{\alpha\beta\gamma}}\equiv
a_{\underline{\alpha\beta\gamma}}+
\sum\limits
_{cycl\atop{(\underline{\alpha}\underline{\beta}\underline{\gamma})}}^{}
\left(b_{\underline{\alpha}\ \underline{\beta}}^{\ \underline{\dot\lambda}}
{\cal P}_{\underline{\gamma\dot\lambda}}+
c_{\underline\alpha}^{\ \underline{\dot\lambda\dot\rho}}
{\cal P}_{\underline{\beta\dot\lambda}}
{\cal P}_{\underline{\gamma\dot\rho}}\right)+
d^{\underline{\dot\lambda\dot\rho\dot\delta}}
{\cal P}_{\underline{\alpha\dot\lambda}}{\cal P}_{\underline{\beta\dot\rho}}
{\cal P}_{\underline{\gamma\dot\delta}}+}\\[0.4cm]
\hspace{9em}{\displaystyle
\mbox{( linear, \,quadratic and cubic terms in
V($\bar V$)})},
\end{array}
\end{equation}
where
$$\begin{array}{c}{\displaystyle d^{\underline{\dot\lambda\dot\rho\dot\delta}}=-8i
\bar M^{-1\underline{\dot\lambda\dot\gamma}}\bar D_{\underline{\dot\gamma}}
\bar M^{-1\underline{\dot\rho\dot\delta}};
c_{\underline\gamma}^{\ \underline{\dot\lambda\dot\rho}}=
4iD_{\underline\gamma}\bar M^{-1\underline{\dot\lambda\dot\rho}}+
4ieF_{\underline{\gamma\dot\delta}}\bar M^{-1\underline{\dot\delta\dot\gamma}}
\bar D_{\underline{\dot\gamma}}\bar M^{-1\underline{\dot\lambda\dot\rho}};}
\\[0.4cm]
{\displaystyle
c^{\underline{\dot\lambda}}{}_{\underline{\alpha(\beta)}}{}^{\underline{\dot\rho}}=
-8e\bar M^{-1\underline{\dot\lambda\dot\gamma}}
\bar M^{-1\underline{\dot\rho\dot\alpha}}
F_{\underline{\alpha(\beta)\dot\alpha},\underline{\dot\gamma}}-4ie
\bar M^{-1\underline{\dot\lambda\dot\gamma}}\bar D_{\underline{\dot\gamma}}\left(
\bar M^{-1\underline{\dot\rho\dot\alpha}}F_{\underline{\alpha(\beta)\dot\alpha}}
\right);}
\\[0.4cm]
{\displaystyle b^{\underline{\dot\lambda}}_{\underline{\alpha\beta}}=
2i\bar M^{-1\underline{\dot\lambda\dot\gamma}}\bar D_{\underline{\dot\gamma}}
M_{\underline{\alpha\beta}}-4e^2\bar M^{-1\underline{\dot\lambda\dot\gamma}}
F_{\underline{\alpha\dot\beta}}\bar M^{-1\underline{\dot\beta\dot\alpha}}
F_{\underline{\beta\dot\alpha},\underline{\dot\gamma}}-4e^2
\bar M^{-1\underline{\dot\lambda\dot\gamma}}
F_{\underline{\alpha\dot\beta},\underline{\dot\gamma}}
\bar M^{-1\underline{\dot\beta\dot\alpha}}\times}\\[0.4cm]
{\displaystyle F_{\underline{\beta\dot\alpha}}-2ie^2
\bar M^{-1\underline{\dot\lambda\dot\gamma}}\bar D_{\underline{\dot\gamma}}
\left(F_{\underline{\alpha\dot\beta}}\bar M^{-1\underline{\dot\beta\dot\alpha}}
F_{\underline{\beta\dot\alpha}}\right);}
\\[0.4cm]{\displaystyle b^{\underline{\dot\lambda}}_{\underline{\alpha(\beta)\gamma}}=
-4e\bar M^{-1\underline{\dot\lambda\dot\alpha}}
F_{\underline{\alpha(\beta)\dot\alpha},\underline{\gamma}}-2ieD_{\underline\gamma}
\left(\bar M^{-1\underline{\dot\lambda\dot\alpha}}  F_{\underline{\alpha(\beta)\dot\alpha}}\right)
-4e^2F_{\underline{\gamma\dot\delta}}\bar M^{-1\underline{\dot\delta\dot\gamma}}
\bar M^{-1\underline{\dot\lambda\dot\alpha}}\times}\\[0.4cm]
{\displaystyle F_{\underline{\alpha(\beta)\dot\alpha},\underline{\dot\gamma}}-2ie^2F_{\underline{\gamma\dot\delta}}
\bar M^{-1\underline{\dot\delta\dot\gamma}}\bar D_{\underline{\dot\gamma}}
\left(\bar M^{-1\underline{\dot\lambda\dot\alpha}}
F_{\underline{\alpha(\beta)\dot\alpha}}\right);}
\\[0.4cm]{\displaystyle a_{\underline{\alpha\beta\gamma}}=-i
D_{\underline\alpha}M_{\underline{\beta\gamma}}+2e^2
F_{\underline{\alpha\dot\beta}}\bar M^{-1\underline{\dot\beta\dot\alpha}}
F_{\underline{\beta\dot\alpha},\underline{\gamma}}+
ie^2D_{\underline\alpha}\left(F_{\underline{\beta\dot\beta}}
\bar M^{-1\underline{\dot\beta\dot\alpha}}F_{\underline{\gamma\dot\alpha}}\right)+}
\\[0.4cm]{\displaystyle 2e^2F_{\underline{\alpha\dot\beta},\underline{\gamma}}
\bar M^{-1\underline{\dot\beta\dot\alpha}}F_{\underline{\beta\dot\alpha}}
-ieF_{\underline{\alpha\dot\delta}}\bar M^{-1\underline{\dot\delta\dot\gamma}}
 \bar D_{\underline{\dot\gamma}}M_{\underline{\beta\gamma}}
+2e^3F_{\underline{\alpha\dot\delta}}
\bar M^{-1\underline{\dot\delta\dot\gamma}}F_{\underline{\beta\dot\beta}}
\bar M^{-1\underline{\dot\beta\dot\alpha}}   F_{\underline{\gamma\dot\alpha},\underline{\dot\gamma}}+}\\[0.4cm]
{\displaystyle 2e^3F_{\underline{\alpha\dot\delta}}
\bar M^{-1\underline{\dot\delta\dot\gamma}}
F_{\underline{\beta\dot\beta},\underline{\dot\gamma}}
\bar M^{-1\underline{\dot\beta\dot\alpha}}F_{\underline{\gamma\dot\alpha}}+
ie^3F_{\underline{\alpha\dot\delta}}
\bar M^{-1\underline{\dot\delta\dot\gamma}}\bar D_{\underline{\dot\gamma}}\left(
F_{\underline{\alpha\dot\beta}}
\bar M^{-1\underline{\dot\beta\dot\alpha}}F_{\underline{\gamma\dot\alpha}}\right).}
\end{array}
$$
${\displaystyle Y_{\underline{\alpha\beta\gamma}}}$ possesses the polynomial structure of
 the third power with respect to ${\displaystyle {\cal P}_{\underline{\alpha\dot\alpha}}}$
 with the coefficient functions depending on the superfield strengths. It is not a function
 of the present constraints, so we are forced to consider it as a new constraint. Again, we
 calculate the
Poisson brackets for ${\displaystyle Y_{\underline{\alpha\beta\gamma}}}$ and
${\displaystyle V^{(1)}_{\underline\delta}}$ to obtain a new fourth power polynomial
 constraint, which should be then considered as a new constraint. In the limit, the above-described procedure leads to an infinite sequence of polynomial constraints of arbitrary high power with respect to
${\displaystyle{\cal P}_{\underline{\alpha\dot\alpha}}}$ with the  coefficient functions
constructed from the superfield strengths. A controlled analysis of the exact form of these infinite constraints chain is rather difficult, since we have not found any recursion procedure for their presentation as functions of  ${\displaystyle Y_{\underline{\alpha\beta}}}$.

The only reason, why we have got infinite set of the constraints, is that  the object ${\displaystyle Y_{\underline{\gamma_1}...\underline{\gamma_n}
\underline{\alpha\beta}}}$  appearing on every stage was considered as the creation of a new
constraint. A possibility to avoid such uncontrolled multiplication of the constraints supposes
 their identical fulfilment
(for the total set of ${\displaystyle{\cal P}_{\underline{\alpha\dot\alpha}}}$) starting from
a certain stage. Taking into account the structure mentioned above, the identical
 fulfilment actually signifies some restriction  for the superfield configurations.
The  identical fulfilment of the
n-th stage constraint yields the identical fulfilment of the next stage constraints.
Although we can't realize this procedure for an arbitrary stage (because of the sophisticated
structure of the appearing  expressions), here we present the explicit consideration of the
first and the second stages.

At the first stage we deal with ${\displaystyle Y_{\underline{\alpha\beta}}}$. For our
 purpose it will be convenient to introduce the SU(2)-decomposition of the superfield strengths
\begin{equation}{\displaystyle
F_{\underline{\alpha\beta}}=-\varepsilon_{\underline{\alpha\beta}}\bar W+
\tau_a^{ij}\tilde F^a_{\alpha\beta}; F_{\underline{\dot\alpha\dot\beta}}=
\varepsilon_{\underline{\dot\alpha\dot\beta}}W+\tau_{aij}\tilde F^a_{\dot\alpha\dot\beta}; F_{\underline{\alpha\dot\alpha}}=\upsilon_{\underline{\alpha\dot\alpha}}+
\tau_a{}^i{}_j\tilde F^a_{\alpha\dot\alpha},}
\end{equation}
and then the matrix inverse to M takes the form
\begin{equation}{\displaystyle
\bar M^{-1\underline{\dot\beta\dot\alpha}}=\frac{i}{2(m-ieW/2)}\left\{
\varepsilon^{\underline{\dot\beta\dot\alpha}}+\sum\limits^\infty_{n=1}\left[
\frac{ie}{2(m-ieW/2)}\right]^n
\tilde F^{\underline{\dot\beta{\dot\alpha}_1}}
\tilde F_{\underline{\dot\alpha}_1\underline{\dot\alpha}_2}\cdots
\tilde F^{\underline{\dot\alpha}_{n-1}\underline{\dot\alpha}_n}\right\}.}
\end{equation}
After substitution of the explicit expressions for the $M$,\, $\bar M$ and N matrices and using the mass shell constraint $\chi={\cal P}^2+m^2\approx0$, one finds:

--the quadratic term\vspace{-0.4cm}
\begin{equation}
\hspace{2.4cm}
\sum\limits^\infty_{n=1}\left[\frac{ie}{2(m-ieW/2)}\right]^n
\tilde F^{\underline{\dot\beta\dot\alpha_1}}\cdots
\tilde F^{\underline{\dot\alpha}_{n-1}\underline{\dot\alpha}_n}=0\ \
\Longrightarrow\ \
\tilde F^{\underline{\dot\beta\dot\alpha}}=0;
\end{equation}

--the linear term
\vspace{-0.5cm}
\begin{equation}
\hspace{11.2cm}
F_{\underline{\alpha\dot\beta}}=0;
\end{equation}

--the free term
\vspace{-0.5cm}
$${\hspace{4cm}\displaystyle -2i(m+ie\bar W/2)\varepsilon_{\underline{\alpha\beta}}+
\frac{2im^2\varepsilon_{\underline{\alpha\beta}}}{(m-ieW/2)}-
e\tilde F_{\underline{\alpha\beta}}=0.}$$
Taking into account the linear independence of the $\tau$-matrices leads to the constraints
\begin{equation}{\displaystyle
\tilde F_{\underline{\alpha\beta}}=0}
\end{equation}
and
$${\displaystyle (m+ie\bar W/2)(m-ieW/2)=m^2}. \eqno (37')$$
After the substitution of  the constraints (35-37) into the superfield Bianchi identities,
 we find the following restrictions for  the physical superfields W and $\bar W$:
 ${\displaystyle
D_{\underline\alpha}\bar W=\bar D_{\underline{\dot\alpha}}W=0}$ and
${\displaystyle D^{ij}W-\bar D^{ij}\bar W=0}$, which isolate $N=2$ Maxwell supermultiplet
 ${\displaystyle(z,\lambda^i_\alpha,C^{ij},v_\mu)}$.
The last constraint $(37')$ is preserved. Its differentiation imposes  additional
restrictions on the chiral superfields W and  $\bar W$ which  eliminate some physical degrees of freedom. Thus, it is impossible to consider ${\displaystyle Y_{\underline{\alpha\beta}}}$
 as an identical constraint.

At the second stage the nullification of the cubic term in ${\displaystyle
{\cal P}_{\underline{\alpha\dot\alpha}}}$ from Eqs.(32) yields the two possibilities:
$$\begin{array}{l}
{\displaystyle a)\ \ \bar D_{\underline{\dot\gamma}}\tilde F_{\underline{\dot\alpha\dot\beta}}=0,\, \bar D_{\underline{\dot\gamma}}W=0;}\\[0.4cm]
{\displaystyle b)\ \ \tilde F_{\underline{\dot\alpha\dot\beta}}=0.}
\end{array}
$$
 The treatment of the quadratic terms leaves the only possibility ${\displaystyle
 \tilde F_{\underline{\dot\alpha\dot\beta}}=0}$ and ${\displaystyle
 \bar D_{\underline{\dot\gamma}}W=0}$ together with the constraint
\begin{equation}{\displaystyle
2\delta^i_jF^{'}_{\alpha\dot\beta,\dot\gamma k}+i\bar D_{\dot\gamma k}
F^i_{\alpha{\dot\beta}j}=0}
\end{equation}
and its complex conjugate, where the spinor-vector superfield strength components were decomposed on
 spin $\frac12$ and spin $\frac32$ parts
\begin{equation}{\displaystyle
F_{\alpha\dot\beta,\dot\gamma k}=\varepsilon_{\dot\beta\dot\gamma}V_{\alpha k}+
F^{'}_{\alpha\dot\beta,\dot\gamma k}.}
\end{equation}
The substution of the expressions (38,39) together with their  complex conjugate into the
 Bianchi identities allows to determine the superfields  V and $\bar V$
$${\displaystyle
V_{\underline\gamma}=\frac{i}{2}D_{\underline\gamma}W, \,                                                         \bar V_{\underline{\dot\gamma}}=-\frac{i}{2}\bar D_{\underline{\dot\gamma}}\bar W.}
$$
 Now, by analogy  with  the first stage, the linear terms impose too strong constraints on the
 chiral superfields W and $\bar W$: ${\displaystyle
D_{\underline\alpha}W=\bar D_{\underline{\dot\alpha}}\bar W=0.}$ And, again, we conclude that
 the identical fulfilment of ${\displaystyle
Y_{\underline{\alpha\beta\gamma}}}$ eliminates physical degrees of freedom.

Although we could not prove it evidently, it is quite reasonable to conjecture that the identical fulfilment of the next stage constraints wlll also eliminate the physical degrees of freedom. Then the necessity for the introduction of nonminimal terms to preserve the k-symmetry becomes
evident.

\begin{center}
{\Large\bf4.  A nonminimal coupling\\ of $N=2$ massive superparticle}
\end{center}

There exists a possibility to introduce some  nonminimal terms for a superparticle possessing not only the electrical charge e, but also an AMM  $\mu$, in such a way that the minimal structure of the interactions caused by the electric charge, will be preserved.
 Taking into account the dimensional reasons
 ($[\mu]=L$  in the system $c=\hbar=1$) we can construct the dimensionless gauge invariant
scalars $\mu F^{\underline{\alpha}}{}_{\underline{\alpha}}$ and
$\mu\bar F_{\underline{\dot\alpha}}{}^{\underline{\dot\alpha}}$ linear in the field
 strenghts. Analogous considerations were used in [19] for the introduction of nonmininal
 terms by means of the extension of the superconnection $1-$form. Then the superparticle action can be written in the form
\begin{equation}
\begin{array}{l}{\displaystyle
S^{(e,\mu)}=-m\int d\tau\sqrt{-F\omega^\mu\omega_\mu}+
m\int d\tau\left(\theta^{\underline\alpha}\dot\theta_{\underline\alpha}+
\bar\theta_{\underline{\dot\alpha}}
\dot{\bar\theta}\mathstrut^{\underline{\dot\alpha}}\right)}
\\[0.4cm]\hspace{3.5em}{\displaystyle
{}+ie\int d\tau\left(\omega^\mu A_\mu+
\dot\theta^{\underline\alpha}A_{\underline\alpha}+
\dot{\bar\theta}\mathstrut_{\underline{\dot\alpha}}
\bar A^{\underline{\dot\alpha}}\right),}
\end{array}
\end{equation}
where ${\displaystyle F=\left(1-\frac{i\mu}{4}F^{\underline\alpha}{}_{\underline\alpha}
\right)\left(1-\frac{i\mu}{4}F_{\underline{\dot\alpha}}{}^{\underline{\dot\alpha}}\right)}$.
The rescaled  mass $m^*=(mF)^{1/2}$ in the first term is  supersymmetric and gauge invariant.
 Similar procedure in the second term would violate global supersymmetry.
 Introduction of the world-line einbein gives
\begin{equation}
\begin{array}{l}{\displaystyle
L^{(e,\mu)}=\frac12\left(\frac{F\omega^2}{g}-gm^2\right)+
m\left(\theta^{\underline\alpha}\dot\theta_{\underline\alpha}+
\bar\theta_{\underline{\dot\alpha}}
\dot{\bar\theta}\mathstrut^{\underline{\dot\alpha}}\right)}
\\[0.6cm]\hspace{3.3em}{\displaystyle
{}+ie\left(\omega^\mu A_\mu+\dot\theta^{\underline\alpha}A_{\underline\alpha}+
\dot{\bar\theta}\mathstrut_{\underline{\dot\alpha}}
A^{\underline{\dot\alpha}}\right)}.
\end{array}
\end{equation}
Note that after the redefinition:$g\rightarrow\frac{g}{F}$
the contribution of AMM can be presented in the form of the potential term $gFm^2$ which
dissapears when $m=0$.

The canonical momentum variables are
\begin{equation}
\begin{array}{c}{\displaystyle
p_\mu=\frac{\partial L}{\partial\dot x}=\frac{F}{g}\omega_\mu+ieA_\mu, p_g=
\frac{\partial L}{\partial\dot g}=0,}\\[0.4cm]
{\displaystyle\pi^i_\alpha=\frac{\partial L}{\partial\dot\theta^\alpha_i}=
-i\frac{F}{g}\omega_{\alpha\dot\alpha}\bar\theta^{\dot\alpha i}-
m\theta^i_\alpha+eA_{\alpha\dot\alpha}\bar\theta^{\dot\alpha i}+
ieA^i_\alpha,}\\[0.4cm]
{\displaystyle \bar\pi_{\dot\alpha i}=
\frac{\partial L}{\dot{\bar\theta}\mathstrut^{\dot\alpha i}}=
-i\frac{F}{g}\theta^\alpha_i\omega_{\alpha\dot\alpha}-m\bar\theta_{\dot\alpha i}+
e\theta^\alpha_iA_{\alpha\dot\alpha}+ie\bar A_{\dot\alpha i}}
\end{array}
\end{equation}
and the corresponding canonical Hamiltonian is given by
\begin{equation}
{\displaystyle H_0=\frac{g}{2F}\left[(p^\mu-ieA^\mu)^2+m^{*2}\right]}
\end{equation}
The definition (42) yields the primary constraints $p_g\approx0$ and
\begin{equation}
{\displaystyle V^i_\alpha=\pi ^i_\alpha+ip_{\alpha\dot\alpha}\bar\theta^{\dot\alpha i}+
m\theta^i_\alpha-ieA^i_\alpha\approx0, \bar V_{\dot\alpha i}=\bar\pi_{\dot\alpha i}+
i\theta^\alpha_i p_{\alpha\dot\alpha}+m\bar\theta_{\dot\alpha i}-
ie\bar A_{\dot\alpha i}\approx0.}
\end{equation}
The total Hamiltonian reads
$${\displaystyle
H=H_0+\lambda^{\underline\alpha}V_{\underline\alpha}+
\bar\lambda_{\dot\alpha}\bar V^{\dot\alpha}+\varphi p_g.}
\eqno (43')
$$
Note that, unlike $H$ and $H_0$, the primary constraints don't "feel" nonminimal terms.
As in Section 3, the temporal conservation of the primary constraints leads to the secondary
constraint ${\displaystyle \chi={\cal P}^2+m^{*2}\approx0}$ and to the linear system of
 equations for Lagrange multipliers coinciding with (23), if the following redefinitions are taken into account
\begin{equation}
{\displaystyle Q_{\underline\alpha}=\frac{eg}{2F}{\cal P}^{\dot\beta\beta}
F_{\beta\dot\beta,\underline\alpha}+\frac{igm^2}{2F}D_{\underline\alpha}F,
\bar Q_{\underline{\dot\alpha}}=\frac{eg}{2F}{\cal P}^{\dot\beta\beta}
F_{\beta\dot\beta,\underline{\dot\alpha}}+\frac{igm^2}{2F}\bar D_{\underline{\dot\alpha}}F.}
\end{equation}
Again, to have the first-class spinorial constraints, generating the k-symmetry, we need to halve the system rank.
The necessary and sufficient condition of the rank halving is the presence of
 the constraints (27,29), as in the section 3. Again the spinorial first-class constraints read
\begin{equation}
{\displaystyle
V^{(1)}_{\underline\alpha}=V_{\underline\alpha}-N_{\underline{\alpha\dot\beta}}
\bar M^{-1\underline{\dot\beta\dot\alpha}}\bar V_{\underline{\dot\alpha}}.}
\end{equation}
So, the  analysis performed in the previous section is also valid here, and we can formulate the conclusion:  the possibility to avoid a sequence of infinite constraints is the identical fulfilment
of the constraint (27). Substituting explicit expressions for M and N matrices and using the mass-shell condition we get:

-the quadratic term\vspace*{-0.4cm}
\begin{equation}\hspace{2.4cm}
\sum\limits^\infty_{n=1}\left[\frac{ie}{2(m-ieW/2)}\right]^n
\tilde F^{\underline{\dot\beta\dot\alpha_1}}
\cdots\tilde F^{\underline{{\dot\alpha}_{n-1}{\dot\alpha}_n}}=0\ \
\Longrightarrow \ \
\tilde F^{\underline{\dot\beta\dot\alpha}}=0;
\end{equation}

-the linear term\vspace*{-0.5cm}
\begin{equation}\hspace{11.5cm}
F_{\underline{\alpha\dot\beta}}=0;                                                                                  \end{equation}

-the free term\vspace{-0.5cm}
$$\hspace{4cm}{\displaystyle -2i(m+ie\bar W/2)\varepsilon_{\underline{\alpha\beta}}+
\frac{2im^2F\varepsilon_{\underline{\alpha\beta}}}{(m-ieW/2)}-
e\tilde F_{\underline{\alpha\beta}}=0.}
$$  in each order.
Taking into account $\tau$-matrices linear independence one has:
\begin{equation}
{\displaystyle \tilde F_{\underline{\alpha\beta}}=0}
\end{equation}
and
$$
{\displaystyle (m+ie\bar W/2)(m-ieW/2)=(m+im\mu\bar W)(m-im\mu W).}
\eqno (49')
$$
Substituting the constraints (47-49) into the Bianchi identities for the superfield strengths
 leads to the standard constraints on $N=2$ physical superfields
\begin{equation}
{\displaystyle
\bar D_{\underline{\dot\alpha}}W=D_{\underline\alpha}\bar W=D^{ij}W-\bar D^{ij}\bar W
=0.}\end{equation}
 The remaining constraint $(49')$ either fixes the AMM magnitude $\mu=\frac{e}{2m}$,
imposing no further constraints on W and $\bar W$, or eliminates the physical degrees
 of freedom.
So, unlike the minimal case, nonminimal one gives rise to the first-class constraints
(and the k-symmetry), but at certain field configurations (50) and $\mu=\frac{e}{2m}$. Then the superparticle Lagrangian is presented as
\begin{equation}
\begin{array}{c}{\displaystyle
L^{(e,\mu(e))}=\frac12\left[\frac{(m-ieW/2)(m+ie\bar W/2)\omega^2}{g}-gm^2\right]}
\\[0.5cm]
{\displaystyle
{}+m\left(\theta^{\underline\alpha}\dot\theta_{\underline\alpha}+
\bar\theta_{\underline{\dot\alpha}}\dot{\bar\theta}\mathstrut^{\underline{\dot\alpha}}
\right)+ie\left(\omega^\mu A_\mu+\dot\theta^{\underline\alpha}A_{\underline\alpha}+
\dot{\bar\theta}\mathstrut_{\underline{\dot\alpha}}\bar A^{\underline{\dot\alpha}}
\right)}
\end{array}
\end{equation}
and the first-class spinorial constraints are written in the form
\begin{equation}
{\displaystyle V^{(1)}_{\underline\alpha}=V_{\underline\alpha}-\frac{i
{\cal P}_{\underline{\alpha\dot\beta}}\bar V^{\underline{\dot\beta}}}{(m-ieW/2)};\ \
\bar V^{(1)}_{\underline{\dot\alpha}}=\bar V_{\underline{\dot\alpha}}-
\frac{i{\cal P}_{\underline{\beta\dot\alpha}}V^{\underline\beta}}{(m+ie\bar W/2)},}
\end{equation}
only half of them being independent:
\begin{equation}
{\displaystyle \frac{i{\cal P}^{\underline{\dot\alpha\alpha}}V^{(1)}_{\underline\alpha}}
{(m+ie\bar W/2)}\approx\bar V^{(1)\underline{\dot\alpha}}.}
\end{equation}
To obtain the explicit expressions for the total Hamiltonian $(43')$ we have to solve
equations (23) subjected to the constraints (47-50) and substitute the solution into $(43')$.
 Equations (23) can be written in the form
\begin{equation}
\begin{array}{c}{\displaystyle
\frac{-ieg}{4F}{\cal P}_{\alpha\dot\beta}\bar D^{\dot\beta i}\bar W+\frac{eg}{4F}
(m+ie\bar W/2)D^i_\alpha W-2i(m+ie\bar W/2)\lambda^i_\alpha+\bar\lambda^{\dot\beta i}
{\cal P}_{\alpha\dot\beta}=0}\\[0.4cm]
{\displaystyle \frac{ieg}{4F}{\cal P}_{\beta\dot\alpha}D^\beta_iW-\frac{eg}{4F}(m-ieW/2)\bar D_{\dot\alpha i}\bar W+2\lambda^\beta_i{\cal P}_{\beta\dot\alpha}-
2i(m-ieW/2)\bar\lambda_{\dot\alpha i}=0,}
\end{array}
\end{equation}
where we used the spinor-vector superfield strength components following from  the Bianchi identities solution
\begin{equation}{\displaystyle
F_{\mu}{}^i_\alpha=\frac{i}{4}(\sigma_\mu\bar D^i)_\alpha\bar W,\ \ F_{\mu\alpha i}=
-\frac{i}{4}(D_i\sigma_\mu)_{\dot\alpha}W.}
\end{equation}
Solving (54) with respect to $\lambda_{\alpha 2}$ and $\bar\lambda_{\dot\alpha 1}$
\footnote{\normalsize As the system rank equals four, we can choose  either
$\lambda_{\alpha 1}$ and $\bar\lambda_{\dot\alpha 2}$ or ($\lambda_{\alpha 2}$ and
 $\bar\lambda_{\dot\alpha 1}$)as independent variables.}, and substituting the solution
in $(43')$  we obtain
\begin{equation}
\begin{array}{c}{\displaystyle
H=\frac{g}{2F}T+\lambda^\alpha_1V^{(1)1}_\alpha+\bar\lambda_{\dot\alpha 2}
\bar V^{(1)\dot\alpha 2}=}\\[0.4cm]
{\displaystyle \frac{g}{2F}\left[\chi-\frac{eV^{\alpha 2}{\cal P}_{\alpha\dot\beta}
\bar D^{\dot\beta}_2\bar W}{4(m+ie\bar W/2)}+\frac{ie}{4}D_{\alpha 2}WV^{\alpha 2}+
\frac{eD^\beta_1W{\cal P}_{\beta\dot\alpha}\bar V^{\dot\alpha 1}}{4(m-ieW/2)}+
\frac{ie}{4}\bar D_{\dot\alpha 1}\bar W\bar V^{\dot\alpha 1}\right]+}\\[0.6cm]
{\displaystyle
\lambda^\alpha_1V^{(1)1}_\alpha+\bar\lambda_{\dot\alpha 2}\bar V^{(1)\dot\alpha 2}
\approx0.}
\end{array}
\end{equation}
Both the spinorial first-class constraints and the bosonic reparametrization generator T are
not SU(2)-invariants, however, the latter may be written in completely invariant fashion
 by means of the shift
\begin{equation}
{\displaystyle\lambda_{\underline\alpha}\longrightarrow\lambda_{\underline\alpha}
+\frac{ieg}{8F}D_{\underline\alpha}W,\ \ \bar\lambda_{\underline{\dot\alpha}}
\longrightarrow\bar\lambda_{\underline{\dot\alpha}}-\frac{ieg}{8F}
\bar D_{\underline{\dot\alpha}}\bar W.}
\end{equation}
Then the full Hamiltonian takes form
\begin{equation}
\begin{array}{c}{\displaystyle
H=\frac{g}{2F}T+\lambda^\alpha_1V^{(1)1}_\alpha+\bar\lambda_{\dot\alpha 2}
\bar V^{(1)\dot\alpha 2}=}\\[0.4cm]
{\displaystyle \frac{g}{2F}\left[({\cal P}^2+m^{*2})-\frac{ie}{4}D^{\underline\alpha}W
V_{\underline\alpha}+\frac{ie}{4}\bar D_{\underline{\dot\alpha}}
\bar V^{\underline{\dot\alpha}}\right]+\lambda^\alpha_1V^{(1)1}_\alpha+
\bar\lambda_{\dot\alpha 2}\bar V^{(1)\dot\alpha 2}.}
\end{array}
\end{equation}
It contains only half of the spinorial Lagrange multipliers and the same number
of the spinorial first-class constraints.

It is straightforward to construct the k-symmetry transformation laws implied by the
action (40):
${\displaystyle
\delta x^\mu=-i\theta_i\sigma^\mu\delta\bar\theta^i+\delta\theta_i\sigma^\mu\bar\theta^i}$,
\begin{equation}
{\displaystyle
{\delta\theta^i_\alpha\choose\delta\bar\theta^{\dot\alpha i}}=
P_I{\kappa^i_\beta(\tau)\choose\bar\kappa^{\dot\beta j}(\tau)}=
\frac12\left({\kappa^i_\alpha-{\displaystyle\frac{im^*\omega_{\alpha\dot\beta}
\bar\kappa^{\dot\beta i}}
{\sqrt{-\omega^2}(m+ie\bar W/2)}}
\atop \bar\kappa^{\dot\alpha i}+
{\displaystyle\frac{im^*\omega^{\dot\alpha\beta}\kappa^i_\beta}
{\sqrt{-\omega^2}(m-ieW/2)}}}\right),}
\end{equation}
where $(\kappa^i_\alpha(\tau),\bar\kappa^{\dot\alpha i}(\tau))$ are arbitrary functions of $\tau$. Here we used the projector [17], which satisfies the well-known conditions $P_I^2=P_I$ and
 $\Gamma^2=1$, to halve the bispinor components
\begin{equation}
P_I=\frac12\delta^i_j(1+\Gamma)=\frac12\delta^i_j\left(
\begin{array}{cc}
\delta^\beta_\alpha &
{\displaystyle\frac{im^*\omega_{\alpha\dot\beta}}{\sqrt{-\omega^2}(m+ie\bar W/2)}}\\
{}&{}\\
{\displaystyle\frac{-im^*\tilde\omega^{\dot\alpha\beta}}{\sqrt{-\omega^2}(m-ieW/2)}} &
\delta_{\dot\beta}^{\dot\alpha}
\end{array}
\right),
\end{equation}
 Introduction of the projector $P_{II}$, with the standard properties $P^2_{I,II}=P_{I,II}$,
 $P_IP_{II}=P_{II}P_I=0$, gives us the way to separate the first- and the second-class
 constraints
\begin{equation}
\begin{array}{c}
{\displaystyle {V^{(1)i}_\alpha\choose\bar V^{(1)\dot\alpha i}}=
P_I{V^j_\beta\choose\bar V^{\dot\beta j}}
=\frac12
\left(V^i_\alpha-
{\displaystyle \frac{i{\cal P}_{\alpha\dot\beta}\bar V^{\dot\beta i}}{(m-ieW/2)}
\sqrt{\frac{m^{*2}}{-{\cal P}^2}}}\atop
\bar V^{\dot\alpha i}+
{\displaystyle \frac{i\tilde{\cal P}^{\dot\alpha\beta}V^i_\beta}{(m+ie\bar W/2)}
\sqrt{\frac{m^{*2}}{{-\cal P}^2}}}\right)}\\[1.0cm]
{\displaystyle{V^{(2)i}_\alpha\choose\bar V^{(2)\dot\alpha i}}=P_{II}{V^j_\beta\choose
\bar V^{\dot\beta j}}=\frac12
\left(V^i_\alpha+{\displaystyle\frac{i{\cal P}_{\alpha\dot\beta}\bar V^{\dot\beta i}}
{(m-ieW/2)}\sqrt{\frac{m^{*2}}{-{\cal P}^2}}}\atop
\bar V^{\dot\alpha i}-{\displaystyle\frac{i\tilde{\cal P}^{\dot\alpha\beta}V^i_\beta}
{(m+ie\bar W/2)}\sqrt{\frac{m^{*2}}{{-\cal P}^2}}}\right)}
\end{array}
\end{equation}
These projectors can be presented in the form
\begin{equation}
{\displaystyle P_{I,II}=\frac12\delta^i_j(1\pm\Gamma)},
\end{equation}
where

$$
{\displaystyle\Gamma=\left(\begin{array}{cc}
0 & {\displaystyle\frac{-i{\cal P}_{\alpha\dot\beta}}{(m-ieW/2)}\sqrt{\frac{m^{*2}}
{-{\cal P}^2}}}\\[0.4cm]
{\displaystyle\frac{i\tilde{\cal P}^{\dot\alpha\beta}}{(m+ie\bar W/2)}\sqrt{\frac{m^{*2}}
{-{\cal P}^2}}} & 0
\end{array}\right)}.
\eqno (62')
$$
The first- and the second-class constraints are (separately) linear dependent:
\begin{equation}
{\displaystyle V^{(1)i}_\alpha\frac{i\tilde{\cal P}^{\dot\beta\alpha}}{(m+ie\bar W/2)}
\sqrt{\frac{m^{*2}}{-{\cal P}^2}}=\bar V^{(1)\dot\beta i}\mbox{,  }
V^{(2)i}_\alpha\frac{-i\tilde{\cal P}^{\dot\beta\alpha}}{(m+ie\bar W/2)}
\sqrt{\frac{m^{*2}}{-{\cal P}^2}}=\bar V^{(2)\dot\beta i}}.
\end{equation}

In the conclusion let us show that the introduced coupling constant $\mu$ actually describes
 the superparticle AMM.
For this purpose we are to consider the term ${\displaystyle\frac{i\mu}{2g}\omega^2(\bar W-W)}$
in (41). Separating the photon part in W superfield component decomposition we
obtain
\begin{equation}
{\displaystyle W=\cdots-2i\theta_i\sigma^{\mu\nu}\theta^iv_{\mu\nu}+\cdots}
\end{equation}
While substituting this expression back into (41) and passing to the pseudoclassical bispinor variables [19]
$${\displaystyle \Psi^i={(-i\omega^2/g)^{\frac12}\theta^i_\alpha\choose
(i\omega^2/g)^{\frac12}\bar\theta^{\dot\alpha i}}}
$$
and introducing the spin operator
 $\Sigma^{\mu\nu}=\frac{i}{4}[\gamma^\mu,\gamma^\nu]$ ($\gamma$-
matrices are taken in the Weyl basis) we find
\begin{equation}
{\displaystyle\left.\frac{i\mu}{2g}\omega^\mu\omega_\mu(\bar W-W)\right|_{\mbox{
\normalsize photon}}=\mu\bar\Psi_i\Sigma^{\mu\nu}\Psi^iv_{\mu\nu}(x)+
\mbox{(higher corrections )}}
\end{equation}
As is seen, the expression (65) is just the ordinary Pauli term.

\begin{center}
{\Large\bf5. Conclusions}
\end{center}

We have examined the $N=2$ extended massive superparticle model [4] (with and without interactions) following to the Dirac prescription and evolving the results [12]. In the free case this model possesses  the k-symmetry which  allows to gauge away a half of fermionic degrees of freedom. Including the minimal interaction with the Abelian gauge superfield ${\displaystyle
A_M(x,\theta,\bar\theta)}$ and demanding the k-symmetry existence, as in the free case,
imposes too strong constraints on the  superfield strengths ${\displaystyle
F_{MN}}$, which eliminate the component fields of the $N=2$ Maxwell multiplet. So we present an explicit proof for the conclusion that the model with minimal coupling actually does not permit k-invariant terms of interaction. To restore the  k-invariance we introduced the nonminimal terms into the Lagrangian of the superparticle, which describes the AMM caused interactions, as we've shown.  Thus the idea of the nonminimal terms introduction, earlier advanced in [12], gets here its explicit realization by means
 of constructing the nonminimal Lagrangian. This Lagrangian is invariant under the k-symmetry transformations, only if the AMM  value of the massive superparticle is rigorously fixed.

\begin{center}
{\Large\bf Acknowledgements}
\end{center}

We would like to thank Ulf Lindstrom and Luca Lusanna for useful discussions. This work was
supported in part by INTAS Grant 93-127-ext and Ukrainian SFFI Grants F4/1751, F5/1794.
 A.Z. is supported by the Grant of Royal Swedish Academy of Sciences.

\begin{center}
{\Large\bf Appendix A}
\end{center}

In the present paper we use following metric signature: ${\displaystyle
\eta_{\mu\nu}=\diag(-,+,+,+)}$. SU(2) and SO(1,3) spinor indices can be raised or lowered in completely covariant way:
$$
\begin{array}{c}{\displaystyle
\theta ^i_\alpha=\varepsilon_{\alpha\beta}\varepsilon^{ij}\theta^\beta_j\mbox{,  }
\theta^\alpha_i=\varepsilon^{\alpha\beta}\varepsilon_{ij}\theta^j_\beta;}\\[0.4cm]
{\displaystyle\bar\theta_{\dot\alpha i}=\varepsilon_{\dot\alpha\dot\beta}
\varepsilon_{ij}\bar\theta^{\dot\beta j}\mbox{,  }\bar\theta^{\dot\alpha i}=
\varepsilon^{\dot\alpha\dot\beta}\varepsilon^{ij}\bar\theta_{\dot\beta j},}
\end{array}
$$
where $\varepsilon^{12}=\varepsilon_{12}=1$.
$\sigma$-matrices have the following properties:
$$
\begin{array}{c}
{\displaystyle\tilde\sigma^{\mu\dot\alpha\alpha}=\varepsilon^{\dot\alpha\dot\beta}
\varepsilon^{\alpha\beta}\sigma^\mu_{\beta\dot\beta}\mbox{,  }
\sigma^\mu_{\alpha\dot\alpha}\tilde\sigma^{\nu\dot\alpha\alpha}=-2\eta^{\mu\nu},}
\\[0.4cm]
{\displaystyle\sigma^\mu_{\alpha\dot\alpha}\tilde\sigma_\mu^{\dot\beta\beta}=
-2\delta_\alpha^\beta\delta^{\dot\beta}_{\dot\alpha}\mbox{,  }
\sigma^{(\mu}_{\alpha\dot\alpha}\tilde\sigma^{\nu)\dot\alpha\beta}=
-2\eta^{\mu\nu}\delta_\alpha^\beta,}\\[0.4cm]
{\displaystyle\tilde\sigma^{(\mu\dot\alpha\alpha}\sigma^{\nu)}_{\alpha\dot\beta}=
-2\eta^{\mu\nu}\delta^{\dot\alpha}_{\dot\beta}\mbox{,  }
\sigma^{[\mu}_{\alpha\dot\alpha}\tilde\sigma^{\nu]\dot\alpha\beta}=
4\sigma^{\mu\nu}{}_\alpha ^\beta,}\\[0.4cm]
{\displaystyle\tilde\sigma^{[\mu\dot\alpha\alpha}\sigma^{\nu]}_{\alpha\dot\beta}=
4\sigma^{\mu\nu}{}^{\dot\alpha}_{\dot\beta}.}
\end{array}
$$
The above formulae can be used to define the scalars:
$$
\begin{array}{c}
{\displaystyle\theta\zeta\equiv\theta^\alpha_i\zeta^i_\alpha=-\zeta\theta\mbox{,  }
\bar\theta\bar\zeta\equiv\bar\theta_{\dot\alpha i}\bar\zeta^{\dot\alpha i}=
-\bar\zeta\bar\theta,}\\[0.4cm]
{\displaystyle\theta\sigma^\mu\bar\zeta\equiv\theta^\alpha_i
\sigma^\mu_{\alpha\dot\alpha}\bar\zeta^{\dot\alpha i}\mbox{,  }
\theta\sigma^{\mu\nu}\zeta\equiv\theta^\alpha_i \sigma^{\mu\nu}{}_\alpha^\beta\zeta^i_\beta.}
\end{array}
$$
The conjugation rules for spinors, derivatives, $\varepsilon$-matrixes and potentials are the following
$$
\begin{array}{c}
{\displaystyle(\theta^\alpha_i)^\dagger=\bar\theta^{\dot\alpha i}\mbox{,  }
(\theta^{\alpha i})^\dagger=-\bar\theta^{\dot\alpha}_i,}\\[0.4cm]
{\displaystyle(D^\alpha_i)^\dagger=\bar D^{\dot\alpha i}\mbox{,  }
(D^{\alpha i})^\dagger=-\bar D^{\dot\alpha i},}\\[0.4cm]
{\displaystyle(A^i_\alpha)^\dagger=\bar A_{\dot\alpha i}\mbox{,  }
(A_{\alpha i})^\dagger=-\bar A^i_{\dot\alpha},}\\[0.4cm]
{\displaystyle
(A^\mu)^\dagger=-A^\mu\mbox{,  }(\vec D F)^\dagger=(-)^F
(\vec D)^\dagger(F)^\dagger,}
\\[0.4cm]
{\displaystyle(\varepsilon_{\alpha\beta})^\dagger=\varepsilon_{\dot\alpha\dot\beta}
\mbox{,  }(\varepsilon^{\alpha\beta})^\dagger=\varepsilon^{\dot\alpha\dot\beta},}
\\[0.4cm]{\displaystyle(\varepsilon_{ij})^\dagger =-\varepsilon^{ij}\mbox{,  }
(\varepsilon^{ij})^\dagger=-\varepsilon_{ij}.}
\end{array}
$$
Bispinor formalism formulae are written as:
$$
{\displaystyle\Psi^i={\xi^i_\alpha\choose\bar\chi^{\dot\alpha i}}\Longrightarrow
\bar\Psi\equiv(\Psi)^\dagger\gamma^0=(-\chi^\alpha_i,\bar\xi_{\dot\alpha i}).}
$$
Dirac matrices in the Weyl basis take form:
$$
{\displaystyle\gamma^\mu=\left(\begin{array}{c}
0\ \ \sigma^\mu\\[0.4cm]
\tilde\sigma^\mu\ \ 0
\end{array}\right)\mbox{;  }\gamma_5\equiv\gamma^0\gamma^1\gamma^2
\gamma^3=i\left(\begin{array}{c}
-I\ \ 0\\[0.4cm]
0\ \ I
\end{array}\right).}
$$
Any vector can be transformed to a bispinor and vice versa:
$$
{\displaystyle v_{\alpha\dot\alpha}=\sigma^\mu_{\alpha\dot\alpha}v_\mu
\mbox{,  }v^\mu=-\frac12\tilde\sigma^{\mu\dot\alpha\alpha}v_{\alpha\dot\alpha}.}
$$

We use the following Poisson brackets definition:
$$
\begin{array}{c}
{\displaystyle\{C,D\}\equiv i\left(\frac{\partial C}{\partial x^\mu}\frac{\partial D}
{\partial p_\mu}+(-)^C\frac{\partial C}{\partial\theta^i_\alpha}
\frac{\partial D}{\partial\pi^\alpha_i}+(-)^C\frac{\partial C}
{\partial\bar\theta^{\dot\alpha i}}\frac{\partial D}{\partial\bar\pi_{\dot\alpha i}}      \right.-}\\
[0.4cm]{\displaystyle\left.(-)^{CD}\frac{\partial D}{\partial x^\mu}\frac{\partial C}
{\partial p_\mu}-(-)^{CD+D}\frac{\partial D}{\partial\theta^i_\alpha}
\frac{\partial C}{\partial\pi^\alpha_i}-(-)^{CD+D}\frac{\partial D}
{\partial\bar\theta^{\dot\alpha i}}\frac{\partial C}{\partial\bar\pi_{\dot\alpha i}}
\right),}
\end{array}
$$
where C(D) grassmannian parity equals to zero, when C(D) carries an even number of the spinor indices and equals to one otherwise.
$$
{\displaystyle\{C,D\}^\dagger\equiv\{D^\dagger,C^\dagger\}.}
$$

\begin{center}
{\Large\bf Appendix B}
\end{center}

This appendix is devoted to the constraints algebra.

a) free superparticle case:
$$
\begin{array}{c}
{\displaystyle\{V^{(1)}_{\underline\alpha},V^{(1)}_{\underline\beta}\}=
\frac{-4i\varepsilon_{\underline{\alpha\beta}}}{(m+\sqrt{-p^2})}\chi\mbox{,  }
\{\bar V^{(1)}_{\underline{\dot\alpha}},\bar V^{(1)}_{\underline{\dot\beta}}\}=
\frac{-4i\varepsilon_{\underline{\dot\alpha\dot\beta}}}{(m+\sqrt{-p^2})}\chi,}\\[0.4cm]
{\displaystyle\{V^{(1)}_{\underline\alpha},\bar V^{(1)}_{\underline{\dot\alpha}}\}=
\frac{-4p_{\underline{\alpha\dot\alpha}}}{(m+\sqrt{-p^2})\sqrt{-p^2}}\chi;}\\[0.4cm]
{\displaystyle\{V^{(2)}_{\underline\alpha},V^{(2)}_{\underline\beta}\}=
4i(m+{\textstyle\sqrt{-p^2}})\varepsilon_{\underline{\alpha\beta}}\mbox{,  }
\{\bar V^{(2)}_{\underline{\dot\alpha}},\bar V^{(2)}_{\underline{\dot\beta}}\}=
-4i(m+{\textstyle\sqrt{-p^2}})\varepsilon_{\underline{\dot\alpha\dot\beta}},}\\[0.4cm]
{\displaystyle\{V^{(2)}_{\underline\alpha},\bar V^{(2)}_{\underline{\dot\alpha}}\}=
4\left(1+\sqrt{\frac{-m^2}{p^2}}\right)p_{\underline{\alpha\dot\alpha}}.}
\end{array}
$$
Each first-class constraint commutes with each second-class one and bosonic first-class constraint $\chi$ commutes with all of them.

b) Superparticle interacting with $N=2$ Maxwell supermultiplet with the fixed AMM
value.
Here the constraints algebra has the following form:
$$
\begin{array}{l}
{\displaystyle\{G_A,G_B\}=\Omega_{AB}+\omega_{1AB}{}^{\underline\alpha}
V_{\underline\alpha}+\omega_{2AB\underline{\dot\alpha}}
\bar V^{\underline{\dot\alpha}}+\omega_{3AB}{}^{\underline{\alpha\beta}}
V_{\underline\alpha}V_{\underline\beta}+}\\[0.4cm]
{\displaystyle\hspace{6em}\omega_{4AB}{}^{\underline\alpha}_{\underline{\dot\beta}}
V_{\underline\alpha}\bar V^{\underline{\dot\beta}}+\omega_{5AB}
{}^{\underline{\dot\alpha\dot\beta}}\bar V_{\underline{\dot\alpha}}
\bar V_{\underline{\dot\beta}},}
\end{array}
\eqno (B.1)
$$
where $G_A=(V_{\underline\alpha}^{(1),(2)},\bar V_{\underline{\dot\alpha}}^{(1),(2)},T)$.
In (B.1) only the first term determines which of the constraints belongs to the first-class and which to the second. That is why we omit below complicated expressions or other terms and use weak equalities
$$
\begin{array}{c}
{\displaystyle\{V^{(1)}_{\underline\alpha},V^{(1)}_{\underline\beta}\}\approx
\frac{-4i\varepsilon_{\underline{\alpha\beta }}}{(m^*+\sqrt{-{\cal P}^2})}
\sqrt{\frac{m+ie\bar W/2}{m-ieW/2}}\chi,}\\[0.6cm]{\displaystyle
\{\bar V^{(1)}_{\underline{\dot\alpha}},\bar V^{(1)}_{\underline{\dot\beta}}\}\approx
\frac{-4i\varepsilon_{\underline{\dot\alpha\dot\beta}}}{(m^*+\sqrt{-{\cal P}^2})}
\sqrt{\frac{m-ieW/2}{m+ie\bar W/2}}\chi,}\\[0.6cm]
{\displaystyle\{V^{(1)}_{\underline\alpha},\bar V^{(1)}_{\underline{\dot\alpha}}\}\approx
\frac{-4{\cal P}_{\underline{\alpha\dot\alpha}}}{(m^*+\sqrt{-{\cal P}^2})\sqrt{-{\cal P}^2}
}\chi;}\\[0.6cm]{\displaystyle
\{V^{(2)}_{\underline\alpha},V^{(2)}_{\underline\beta}\}\approx
-4i(m^*+\sqrt{-{\cal P}^2})\sqrt{\frac{m+ie\bar W/2}{m-ieW/2}}
\varepsilon_{\underline{\alpha\beta}},}\\[0.6cm]
{\displaystyle\{\bar V^{(2)}_{\underline{\dot\alpha}},
\bar V^{(2)}_{\underline{\dot\beta}}\}\approx
-4i(m^*+\sqrt{-{\cal P}^2})\sqrt{\frac{m-ieW/2}{m+ie\bar W/2}}
\varepsilon_{\underline{\dot\alpha\dot\beta}},}\\[0.6cm]{\displaystyle\{V^{(2)}_{\underline\alpha},
\bar V^{(2)}_{\underline{\dot\alpha}}\}\approx
4{\cal P}_{\underline{\alpha\dot\alpha}}\left(1+\sqrt{\frac{-m^{*2}}{{\cal P}^2}}\right);}
\\[0.6cm]
{\displaystyle\{V^{(1)}_{\underline\alpha},V^{(2)}_{\underline\beta}
(\bar V^{(2)}_{\underline{\dot\beta}})\}\approx0\mbox{,  }
\{\bar V^{(1)}_{\underline{\dot\alpha}},V^{(2)}_{\underline\beta}
(\bar V^{(2)}_{\underline{\dot\beta}})\}\approx0\mbox{,  }
\{T,V^{(1),(2)}_{\underline\alpha}(\bar V^{(1),(2)}_{\underline{\dot\alpha}})\}
\approx0.}
\end{array}
$$

\begin{center}
{\Large\bf References}
\end{center}
1. J.Hughes, J.Liu and J.Polchinski, Phys.Lett. {\bf180B} (1986) 370;\\
\hspace*{0.8em} E.Bergshoeff, E.Sezgin and P.K.Townsend, Phys.Lett. {\bf189B} (1987) 75;\\
\hspace*{0.5cm}Ann.Phys. {\bf185} (1988) 330.\\
2. J.Polchinski, T.A.S.I. lectures on D-branes, hep-th/9611050;\\
\hspace*{0.8em} M.Duff, Supermembranes, hep-th/9611203;\\
\hspace*{0.8em} P.K.Townsend, Four lectures on M-theory, hep-th/9612121.\\
3. W.Siegel, Phys.Lett.{\bf128B} (1983) 397.\\
4. J.A.de Azc\'arraga and J.Lukierski, Phis.Lett. {\bf113B} (1982) 170;\\
\hspace*{0.8em} Phys.Rev {\bf D28} (1983) 1337.\\
5. M.B.Green, J.H.Schwarz and E.Witten, Superstring theory (C.U.P. Cam-\\                                       \hspace*{0.5cm} bridge, 1987);\\ \hspace*{0.8em}
L.Brink and J.H.Schwarz, Phys.Lett. {\bf100B} (1981) 310;\\
6. J.A.de Azc\'arraga and J.Lukierski, Phys.Rev. {\bf D38} (1988) 509.\\
7. J.A.Shapiro and C.C.Taylor, Phys.Rep. {\bf191} (1990) 221;\\
\hspace*{0.8em} G.V.Grigoryan, R.P.Grigoryan and I.V.Tyutin, Teor.Mat.Fiz. {\bf111} (1997) \hspace*{0.5cm}389.\\
8. A.Yu.Nurmagambetov, J.J.Rosales and V.I.Tkach, JETP Lett. {\bf60} (1994) \hspace*{0.5cm}145;\\
\hspace*{0.8em} I.A.Bandos and A.Yu.Nurmagambetov, Class. Quant. Grav. {\bf14} (1997) \hspace*{0.5cm}1597.\\
9. A.A.Zheltukhin and D.V.Uvarov, JETP Lett. {\bf67} (1998) 888.\\
10. I.A.Bandos and A.A.Zheltukhin, Fortschr. der Phys. {\bf41} (1993) 619.\\
11. J.Wess and J.Bagger, Supersymmetry and supergravity (P.U.P., Prince-\\
\hspace*{0.9cm}ton, 1983).\\
12. L.Lusanna and B.Milevski, Nucl.Phys. {\bf B247} (1984) 396.\\
13. A.Barducci, Phys.Lett. {\bf118B} (1982) 112.\\
14. A.Barducci, R.Casalbuoni and L.Lusanna, Nuovo Cim. {\bf35A} (1976) 377.\\
15. A.A.Zheltukhin, Teor.Mat.Fiz. {\bf65} (1985) 151.\\
16. P.A.M.Dirac, Lectures on quantum mechanics (Academic, New York,\\
\hspace*{0.9cm}1964).\\
17. J.M.Evans, Nucl.Phys. {\bf B331} (1990) 711.\\
18. P.West, Introduction to supersymmetry and supergravity (World Scienti-\\
\hspace*{0.9cm}fic, 1986).\\
19. A.A.Zheltukhin and V.V.Tugay, JETP Lett. {\bf61} (1995) 532;\\
\hspace*{0.9cm}Yad.Fiz.{\bf61} (1998) 325;
 hep-th/9706114.
\end{document}